\documentclass[12pt]{article}
\usepackage{graphics}
\usepackage{psfrag}
\usepackage{epsfig}
\begin{document}
\title{{Quark energy loss  and shadowing in nuclear Drell-Yan process }
}
\author{ Chun-Gui Duan $^{1,2,5}$\thanks{E-mail:duancg$@$mail.hebtu.edu.cn}
Shu-Wen Cui $^3$  \\
Zhan-Yuan Yan  $^{4}$ Guang-Lie Li $^{2,5}$
 \\
{\small 1.Department of Physics, Hebei Normal University, Shijiazhuang ,050016,China}\\
{\small 2.Institute of high energy physics ,The Chinese academy of sciences ,Beijing,100039,China}\\
{\small 3.Physics department,Cangzhou Teachers' College,Cangzhou,061000,China}\\
{\small 4.Department of Applied Physics,North China Electric Power University,Baoding,071003,China}\\
{\small
5.CCAST(WorldLaboratory).P.O.Box8730,Beijing,100080,China}}

\date{}
\maketitle
\baselineskip 9mm
\vskip 0.5cm
\begin{center}
\begin{minipage}{134mm}
\begin{center}
Abstract
\end{center}
The energy loss effect in  nuclear matter is another nuclear
effect apart from the nuclear effects on the parton distribution
as in deep inelastic scattering process. The quark energy loss can
be measured best by the nuclear dependence of the high energy
nuclear Drell-Yan process.   By means of three kinds of  quark
energy loss parameterizations given  in literature and  the
nuclear parton distribution extracted only with lepton-nucleus
deep inelastic scattering experimental data, measured Drell-Yan
production cross sections are analyzed for 800GeV proton incident
on a variety of nuclear targets from FNAL E866. It is shown that
our results with considering the energy loss effect are much
different from these of the FNAL E866 who analysis the
experimental data with the nuclear parton distribution functions
obtained by using the deep inelastic lA collisions and pA nuclear
Drell-Yan data . Considering the existence of energy loss effect
in Drell-Yan lepton pairs production,we suggest that the
extraction of nuclear parton distribution functions should not
include Drell-Yan experimental data.


\end{minipage}
\end{center}

\vskip 0.5cm

\hspace{0.5cm}Understanding the initial stages of
ultrarelativistic heavy ion collisions is of utmost importance in
order to understand the outcome of the high energy heavy ion
experiments, such as the BNL relativistic heavy ion collider(RHIC)
and CERN large hadron collider (LHC). Understanding the
modifications of the parton distribution functions  and the parton
energy loss in nuclei should be the first important step towards
pinning down the initial conditions of a heavy-ion collision and
understanding of $J/\psi$ production  which is required if it is
to be used as a signal for the quark-gluon plasma in relativistic
heavy ion collisions.

\hspace{0.5cm}The high energy inelastic hadron-nucleus collisions
has been studied for many decades by both the nuclear and particle
physics communities $^{[1]}$. By means of the nuclei, we can study
the space-time development of the strong interaction during its
early stages, which is inaccessible between individual hadrons. In
high energy inelastic hadron-nucleus scattering, the projectile
rarely retains a major fraction of its momentum after traversing
the nucleus. Rather, its momentum is shared by several produced
particles,which form a hadron jet in the forward direction.The
classical description of this phenomena is that the  projectile
suffers multiple collisions and repeated energy loss in the
nuclear matter. In other words,each quark or gluon in the
projectile can loss a finite fraction of its energy  in the
nuclear target due to QCD bremsstrahlung$^{[2]}$. The Drell-Yan
reaction$^{[3]}$ on nuclear targets provides, in particular, the
possibility of probing the propagation of quark through nuclear
matter, with the produced lepton pair carrying away the desired
information on the projectile quark after it has travelled in the
nucleus. Only initial-state interactions are important in
Drell-Yan process since the dimuon in the final state does not
interact strongly with the partons in the nuclei. This makes
Drell-Yan scattering an ideal tool to study energy loss$^{[4]}$.
In addition,Drell-Yan reaction is closely related to DIS of
leptons , but unlike DIS ,it is directly sensitive to antiquark
contributions in target parton distributions. When DIS on nuclei
occurs at $x<0.08$, where x is the parton momentum fraction, the
cross section per nucleon decreases with increasing nucleon number
A due to shadowing$^{[5]}$. Shadowing should also occur in
Drell-Yan dimuon production at small $x_2$, the momentum fraction
of the target parton, and theoretical calculations indicate that
shadowing in the DIS and Drell-Yan reactions has a common origin
$^{[6]}$.Therefore,shadowing and initial state partonic energy
loss are processes that occur in  the proton-induced Drell-Yan
reaction on nuclei.

\hspace{0.5cm}In order to describe the modification of the initial
state parton distributions in nucleus , a  variety of approach to
this question exist in the literature$^{[7]}$. Recently,there were
two trials without model dependence to obtain nuclear parton
distributions  from  the existing world experimental data. In
1999, Eskola, Kolhinen, Ruuskanen  and  Salgado(EKRS)$^{[8]}$
suggested a set of nuclear parton distributions ,which are studied
within a framework of the DGLAP evolution . The measurements of
$F^A_2/F^D_2$ in deep inelastic $lA$ collisions, and Drell-Yan
dilepton cross sections measured in $pA$ collisions were used as
constraints.The kinematical ranges are $10^{-6}\leq x\leq 1$  and
$2.25GeV^2\leq Q^2\leq10^{4}GeV^{2}$  for nuclei from deuteron to
heavy ones .With the nuclear parton distributions , the calculated
results agreed very well with the relative EMC and Fermilab E772
experimental data$^{[9]}$ . In 2001, Hirai,Komano and
Miyama(HKM)$^{[10]}$ proposed two types  of nuclear parton
distributions which were obtained by quadratic and cubic type
analysis, and determined by a $\chi^2 $ global analysis of
existing experimental data on nuclear structure functions without
including the proton-nucleus Drell-Yan process .The kinematical
ranges covered $10^{-9}\leq x\leq 1$  and
 $1GeV^2\leq Q^2\leq10^{5}GeV^{2}$  for nuclei from deuteron to heavy
ones . As a result, they obtained reasonable fit to the measured
experimental data of $F_2$.In this work , we will use both
parameterizations and investigate the nuclear dependence of the
Drell-Yan process.

\hspace{0.5cm}In order to research the partonic energy loss in
nuclei, Fermilab Experiment866(E866) $^{[11]}$ performed the
precise measurement of the ratios of the Drell-Yan cross section
per nucleon for an 800GeV proton beam incident on Be, Fe and W
target at larger values of $x_1$,the momentum fraction of the beam
parton, larger values of $x_F$($\approx x_1-x_2$),and smaller
values of $x_2$ than reached by the previous experiment, Fermilab
E772$^{[9]}$. The extended kinematic coverage of E866
significantly increases its sensitivity to energy loss and
shadowing.This is the first experiment on the energy loss of quark
passing through a cold nucleus.

\hspace{0.5cm}After the E866 experimental data was
reported,several groups have given their theoretical analysis of
the data$^{[12,13,14]}$. In previous report$^{[14]}$ , by means of
EKRS and HKM nuclear parton distribution functions, we
investigated the Drell-Yan production cross section ratios for
800GeV proton incident on a variety of nuclear targets  in the
framework of  Glauber model. We found that the theoretical results
with energy loss are in good agreement with  the Fermilab E866
experiment by means of HKM nuclear parton distributions. However,
the calculated results without energy loss can give good fits by
using  EKRS nuclear parton distribution functions. In this letter,
the nuclear dependence of the pA Drell-Yan production cross
sections are studied  by combining the quark energy loss
parametrization$^{[15,18,19]}$ given  in literature with the EKRS
and HKM nuclear parton distribution.

\hspace{0.5cm}In the Drell-Yan process,the leading-order
contribution is quark-antiquark annihilation into a lepton pair.
The annihilation cross section can be obtained from the
$e^{+}e^{-}\rightarrow\mu^{+}\mu^{-}$ cross section by including
the color factor $\frac{1}{3}$ with the charge $e^{2}_{f}$ for the
quark of flavor $f$.
\begin{equation}
   \frac{d\hat{\sigma}}{dM}=\frac{8\pi\alpha^2}{9M}e^2_f\delta(\hat{s}-M^2),
\end{equation}
where $\sqrt{\hat{s}}=(x_1x_2s)^{1/2}$,is the center of mass
system (c.m.system) energy of $q\bar{q}$ collision
,$x_1$(resp.$x_2$)is the momentum fraction carried by the
projectile (resp.target) parton, $\sqrt{s}$ is the center of mass
energy of the hadronic collision, and $M$ is the invariant mass of
the produced dimuon. The hadronic Drell-Yan differential cross
section is then obtained from the convolution of the above
partonic cross section with the quark distributions in the beam
and in the target :
\begin{equation}
 \frac{d^2\sigma}{dx_1dM}=K\frac{8\pi\alpha^2}{9M}\frac{1}{x_1s}
 \sum_{f}e^2_f[q^p_f(x_1)\bar{q}^A_f(x_2)
 +\bar{q}^p_f(x_1)q^A_f(x_2)],
\end{equation}
where $ K$ is the high-order QCD correction, $\alpha$ is the
fine-structure constant, the sum is carried out over the light
flavor $f=u,d,s$, and $q^{p(A)}_{f}(x)$ and ${\bar
q}^{p(A)}_{f}(x)$ are the quark and anti-quark distributions in
the proton (nucleon in the nucleus A). In order to obtain the
$x_1$ dependence of Drell-Yan production, we shall  deal in the
following with the single differential cross section,
\begin{equation}
 \frac{d\sigma}{dx_1}=K\frac{8\pi\alpha^2}{9x_1s}
 \sum_{f}e^2_f\int\frac{dM}{M}[q^p_f(x_1)\bar{q}^A_f(x_2)
 +\bar{q}^p_f(x_1)q^A_f(x_2)],
\end{equation}
where the integration over the dimuon mass is performed in the
range given from E866 experiment.

\hspace{0.5cm}Now let us take into account of the energy loss of
the fast quarks moving through the cold nuclei.There were three
expressions for the average change  in incident parton momentum
fraction in literature .The first was given by Gavin and
Milana$^{[15]}$ in 1992.They analyzed the depletion at high
Feynman x in large relative to small nuclear targets from
measurements of the Drell-Yan process and of charmonium and
bottomonium production at CERN$^{[17]}$ and Fermilab
E772$^{[9,16]}$ . They proposed  that energy loss due to multiple
parton scattering can consistently explain the high-$x_F$
depletion in the Drell-Yan process and $J/\psi$ production.The
following expression was assumed for the average change  in
incident parton momentum fraction :
\begin{equation}
\Delta x_1= {\kappa}_1x_1A^{1/3},
\end{equation}
where the factor ${\kappa}_1$ may have a $Q^2$ dependence.From
comparison to the experimental data and neglecting shadowing, an
upper bound  on $dE/dz$--the incident quark energy loss per unit
length --- was obtained,the value is $1.5GeV/fm$.This result was
questioned by Brodsky and Hoyer$^{[18]}$,who argued that the time
scale for gluon bremsstrahlung need to be taken into account.By
analogy with the photon bremsstrahlung process , they derived the
quantum mechanical bound on the amount of radiative energy loss
suffered by high energy quarks and gluons in nuclear matter,
\begin{equation}
\Delta x_1 \approx \frac{{\kappa}_2}{s}A^{1/3},
\end{equation}
where  $s$ denotes  the square of the proton-nucleus center of
mass energy. The formulation developed by Brodsky and Hoyer was
extended by Baier et al.$^{[19]}$ They studied the medium-induced
$p_{\perp}$-broadening and induced gluon radiation spectrum of a
high energy quark or gluon traversing a large nucleus.Multiple
scattering of the high energy parton in nuclei was treated in the
Glauber approximation.  Baier et al. found that the partonic
energy loss  depended on a characteristic length and the
broadening of the squared transverse momentum of the parton. For
finite nuclei,both factors vary as $A^{1/3}$, so  they predicted
\begin{equation}
\Delta x_1 \approx \frac{{\kappa}_3}{s}A^{2/3}.
\end{equation}
Using these energy loss expressions , we can obtain empirical
values for the $\kappa$'s by fitting  the FNAL E866 Drell-Yan
cross section ratios versus the incident proton's momentum
fraction and dimuon effective mass.

\hspace{0.5cm}If considering the  quark energy loss in nuclei, the
incident quark momentum fraction can be shifted from
$x'_1=x_1+\Delta x_1$ to $x_1$ at the point of fusion. Combining
the shadowing with initial state energy loss ,the production cross
section in pA Drell-Yan process can be written as
\begin{equation}
 \frac{d\sigma}{dx_1}=K\frac{8\pi\alpha^2}{9x_1s}
 \sum_{f}e^2_f\int\frac{dM}{M}[q^p_f(x'_1)\bar{q}^A_f(x_2)
 +\bar{q}^p_f(x'_1)q^A_f(x_2)].
\end{equation}
In order to pin down quark energy loss by comparing with the
experimental data from E866 collaboration$^{[11]}$,we introduce
the nuclear Drell-Yan ratios as:
\begin{equation}
R_{A_{1}/A_{2}}(x_{1})=\frac{d\sigma^{p-A_{1}}}{dx_ {1}}/{\frac
{d\sigma^{p-A_{2}}}{dx_{1}}}.
\end{equation}
The integral range on M is determined according to the E866
experimental kinematic region.In our theoretical analysis
,$\chi^2$ is calculated with the Drell-Yan differential cross
section rations $R_{A_1/A_2}$ as
\begin{equation}
\chi^2=\sum\limits_{j}\frac{(R^{data}_{A_1/A_2,j}-R^{theo}_{A_1/A_2,j})^2}
{(R^{err}_{A_1/A_2,j})^2},
\end{equation}
where the experimental error is given by systematic errors as
$R^{err}_{A_1/A_2,j} $, and $ R^{data}_{A_1/A_2,j}$(
$R^{theo}_{A_1/A_2,j}$) indicates the experimental data
(theoretical values ) for the ratio$R_{A_{1}/A_{2}}$.

\hspace{0.5cm} Taking advantage of the EKRS$^{[8]}$ nuclear parton
distribution functions and
$\kappa_1=0,\kappa_2=0,\kappa_3=0$(without energy loss
effects),the obtained $\chi^2$ value is $\chi^2=51.4$  for the 56
total data points. The $\chi^2$ per degrees of freedom is given by
$\chi^2/d.o.f.=0.918$. It is apparent that theoretical results
without energy loss effects agree very well with the E866
experimental data ,which results from EKRS parametrization of
nuclear parton distributions studied with including the E772
Drell-Yan data $^{[16]}$ . Because of employing the EKRS global
phenomenological analysis, the E866 analysis missed the energy
loss effect of quarks in nuclei.

\hspace{0.5cm}In what follows,we consider also combining HKM cubic
type of nuclear parton distribution$^{[10]}$  with the three quark
energy loss parameterizations. With
$\kappa_1=0,\kappa_2=0,\kappa_3=0$(without energy loss
effects),the obtained  $\chi^2$ per degrees of freedom is
$\chi^2/d.o.f.=2.516$. With
$\kappa_1=0.0036,\kappa_2=3.45,\kappa_3=0.46$(with energy loss
effects),the obtained  $\chi^2$ per degrees of freedom are
$\chi^2/d.o.f.=1.00048, 1.022, 1.053$, respectively.  The results
given by HKM quadratic type  are nearly the same as these above.
As an example,the calculated results with energy loss expression
 Eq.(5)are shown in Fig.1 and Fig.2. which is the Drell-Yan cross
section ratios for Fe  to Be and   W to Be as functions of $x_1$
for various interval of $M$,respectively.The solid curves are the
ratios with only the nuclear  effect on the parton distribution as
in DIS scattering process, and the dotted curves correspond to an
energy loss effect: $\kappa_2=3.45 GeV^2$ with nuclear effect on
structure function .From comparison with the experimental data ,
it is found that our theoretical results with energy loss effect
are in good agreement with the Fermilab E866.However,the values of
$\kappa_1,\kappa_2,\kappa_3$ are  very different from those by the
E866 Collaboration$^{[11]}$,which results from employing EKRS
nuclear parton distributions. The EKRS analysis itself included
the Drell-Yan data from E772,with the presumption that the
low-$x_2$ nuclear dependence arose entirely from shadowing.This
clearly introduced an inconsistency into the E866 search for
energy loss. Therefore,we can conclude that the extraction of
nuclear parton distribution functions should not include Drell-Yan
experimental data because of the energy loss of the incoming
partons.

\hspace{0.5cm}In Eq.(4), the average loss of projectile parton
vary with the variable $x_1$. It is difficult to obtain a constant
energy loss $dE/dz(GeV/fm)$.$\kappa_3=0.46$ shows  us that the
observed energy loss of the incident quarks is $\Delta E<0.21
GeV/fm^2\times L^2$.$\kappa_2=3.45$ in Eq.(5)indicates that the
incident quarks lose energy with a constant rate
$dE/dz=2.04GeV/fm$ if $<L>=3/4(1.2A^{1/3})fm$ is used as the
average path length. The value of $dE/dz$ is not  obviously
consistent with that given by M.B.Johnson et al.$^{[12]}$,which is
$-dE/dx=3.14\pm 0.53 GeV/fm$.M.B.Johnson et al. examined the
effect of initial state energy loss on the Drell-Yan cross section
ratios from E866 by employing a new formulation of the Drell-Yan
process in the rest frame of nucleus.It is obvious that an
intuitive explanation of the quark energy loss depends on the
reference frame of the target nucleus. It is well-known that the
physical observables can not depend on our choice of reference
frame. Therefore,further theoretical and experimental studies are
needed to determine the value of incident quark energy loss per
unit length,and verify whether the energy loss is linear or
quadratic with the path length.

\hspace{0.5cm}With considering the existence of energy loss effect
in Drell-Yan lepton pairs production,we suggest that the
extraction of nuclear parton distribution functions should not
employ pA  Drell-Yan experimental data.Although there are
currently abundant data on electron and moun deep inelastic
scattering, it is difficult to determine valence quark
distributions in the small x region and the anti-quark
distributions .Only valence quark distributions in medium x region
can be relatively well determined. It is well considered that the
precise nuclear parton distributions must be known in order to
calculate cross sections of high energy nuclear reactions
accurately and find a signature of quark-gluon plasma in high
energy heavy-ion reactions.We suggest using precise neutrino
scattering experimental data ,which can provide a good  method for
measuring the $F_2(x,Q^2)$ and $xF_3(x,Q^2)$ structure
functions.Using the average of $xF_3^{\nu A}(x,Q^2)$and
$xF_3^{\bar{\nu}A}(x,Q^2)$ , the valence quark distribution
functions can be well pinned down$^{[20]}$. Combining the lepton
inelastic scattering data with the neutrino scattering
experiments, valence quark and anti-quark distribution functions
will be obtained in the future,which makes us good understanding
the energy loss effect in high energy nuclear collisions.

\hspace{0.5cm}In summary, we have made a leading-order analysis of
E866 data in nuclei  by taking into account of the energy loss
effect of fast quarks . Our theoretical results with quark energy
loss are in good agreement with the Fermilab E866 experiment by
means of the parametrization of nuclear parton distributions
studied without nuclear Drell-Yan process, which is the same as
that in our previous work$^{[14]}$.we suggest that the extraction
of nuclear parton distribution functions should not employ pA
Drell-Yan experimental data.In order to determine the amount of
energy loss and the appropriate path length,we desire to  operate
precise measurements of the nuclear dependent of Drell-Yan
production at the Fermilab Main Injector(FMI,120GeV proton
beam)$^{[21]}$ and the Japan Hadron Facility(JHF,50GeV proton
beam)$^{[22]}$,where shadowing effect disappears and energy loss
effect of fast quarks could provide the dominant nuclear
dependence.

{\bf Acknowledgement:}This work is partially supported by  Natural
Science Foundation of China(90103020,10075057,10175074) , CAS
Knowledge Innovation Project (KJCX2-SW-N02),Major State Basic
Research Development Program (G20000774),  Natural Science
Foundation of Hebei Province(103143)

\vskip 2cm
\begin{center} Figure caption
\end{center}
Fig.1 The nuclear Drell-Yan cross section ratios
$R_{{A_1}/{A_2}}(x_1)$ on Fe to Be for various intervals M. Solid
curves correspond to nuclear effect on structure function  .
Dotted curves  show the combination of shadowing and energy loss
effect  with HKM cubic type of nuclear parton distributions .The
experimental data are taken from the E866[11].

Fig.2 The nuclear Drell-Yan cross section ratios
$R_{{A_1}/{A_2}}(x_1)$ on W to Be for various intervals M. The
comments is the same as Fig.1

\newpage
\begin{figure}
\centering
\includegraphics{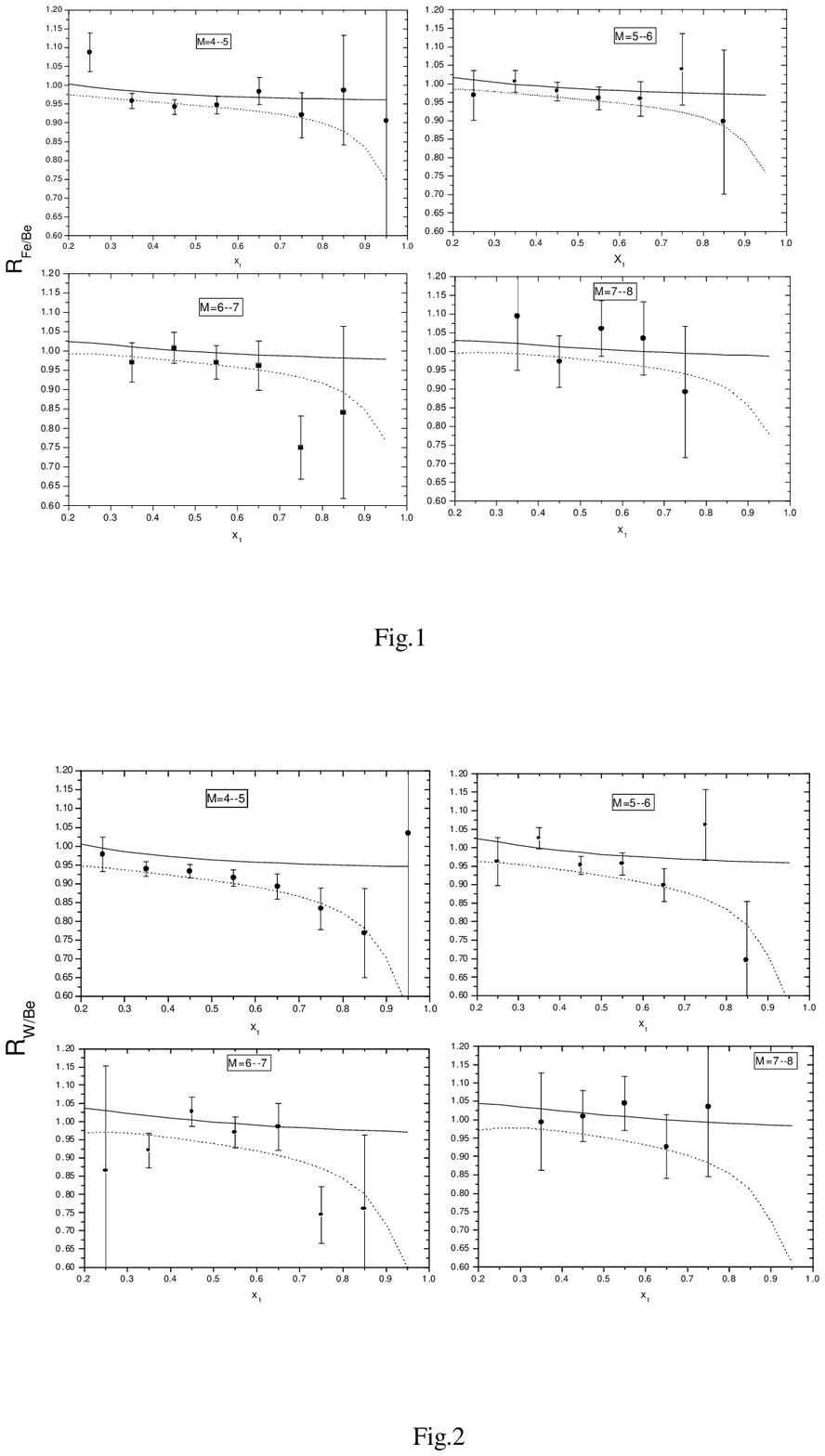}
\end{figure}



\end{document}